\documentclass[10pt]{amsart}

\usepackage{amsmath, amssymb}

\newcommand{\Pin}{\mathrm{Pin}}

\newcommand{\OO}{\mathrm{O}(3,1)}

\newcommand{\R}{\mathbb{R}}
\newcommand{\cR}{\mathcal{R}}
\newcommand{\cE}{\Sigma}
\newcommand{\C}{\mathbb{C}}

\title{Pin Groups in General Relativity}
\author[Bas Janssens]{Bas Janssens}
\address{TU Delft\\
Mekelweg 4, 2628 CD Delft, The Netherlands}
\email{B.Janssens@tudelft.nl}

\begin{document}

\begin{abstract}
There are eight possible Pin groups that can be used to describe
the transformation behaviour of fermions under parity and time reversal.
We show that
only \emph{two} of these are
compatible with general relativity, in the sense that the
configuration space of fermions coupled to gravity
transforms appropriately under the space-time diffeomorphism group.
\end{abstract}

\maketitle

\section{Introduction}\label{sec:Introduction}

For bosons, the space-time transformation behaviour is governed by the Lorentz group
$\OO$, which comprises four connected components.
Rotations and boosts are contained in the connected component of unity,
the proper ortho\-chro\-nous Lorentz group $\mathrm{SO}^{\uparrow}(3,1)$.
Parity ($P$) and time reversal ($T$)
are encoded in the other three connected components of the Lorentz group,
the translates of
$\mathrm{SO}^{\uparrow}(3,1)$
by $P$, $T$ and $PT$.

For fermions, the space-time transformation behaviour is governed by a
double cover of $\mathrm{O}(3,1)$.
Rotations and boosts are described by
the unique simply connected double cover of
$\mathrm{SO}^{\uparrow}(3,1)$, the spin group $\mathrm{Spin}^{\uparrow}(3,1)$.
However,
in order to account for parity and time reversal,
one needs to extend this cover from
$\mathrm{SO}^{\uparrow}(3,1)$ to the full Lorentz group $\mathrm{O}(3,1)$.

This extension is by no means unique. There are no less than \emph{eight}
distinct double covers of $\mathrm{O}(3,1)$ that agree with
$\mathrm{Spin}^{\uparrow}(3,1)$ over $\mathrm{SO}^{\uparrow}(3,1)$.
They are the Pin groups $\mathrm{Pin}^{abc}$,
characterised by the
property that the elements $\Lambda_{P}$ and $\Lambda_{T}$
covering $P$ and $T$ satisfy $\Lambda_{P}^2 = -a$,
$\Lambda_{T}^2 = b$ and $(\Lambda_{P}\Lambda_{T})^2 = -c$,
where $a$, $b$ and $c$ are either $1$ or $-1$ (cf.~\cite{Da88,Ch94}).

In this paper, we show that the consistent description of fermions
in the presence of General Relativity (GR) imposes severe restrictions
on the choice of Pin group. In fact, we find that only \emph{two} of the eight Pin groups
are admissible: the group
$\Pin^+ = \Pin^{++-}$ and the group
$\Pin^- = \Pin^{---}$.
The source of these restrictions is the double cover
of the frame bundle, which, in the context of GR, is needed
in order to obtain
an infinitesimal action of the space-time diffeomorphism group on the configuration space of
fermions coupled to gravity.

We derive these restrictions in the `universal spinor bundle approach' for fermions coupled to
gravity, as developed in \cite{Ko72,BG92,AWW16} for the Riemannian and in
\cite{BaerGauduchonMoroianu2005,J16,J17,MuellerNowaczyk2017} for the Lorentzian
case. However, our results remain valid in other formulations that are covariant under infinitesimal space-time diffeomorphisms, such as the `global'
approach of \cite{DP86,Da88,Swift93,DD13}. To underline this point, we highlight the role of the space-time diffeomorphism group
in restricting the admissible Pin groups.

Selecting the correct Pin groups is important
from a fundamental point of view -- it determines the transformation
behaviour of fermionic fields under reflections --
but also because
the Pin group can affect
observable quantities such as currents \cite{CWM88,WM90,W03}.
Due to their transparent definition in terms of Clifford
algebras, the `Cliffordian' Pin groups
$\Pin(3,1) = \Pin^{+-+}$
and $\Pin(1,3) = \Pin^{-++}$
have attracted much attention \cite{GS87,CWM88,LM89,St94,BWGK01}.
Remarkably, the two Pin groups $\Pin^+$ and
$\Pin^-$
that are compatible with GR are
\emph{not} the widely used Cliffordian Pin groups
$\Pin(3,1)$
and $\Pin(1,3)$.

\section{The lorentzian metric}

In order to establish notation, we briefly recall
the \emph{frame} or \emph{vierbein formalism}
for a Lorentzian metric $g$ on a
four-dimensional space-time manifold~$M$.

A \emph{frame} $e_{x}$
based at
$x$ is a basis $e^{\mu}_{a}\partial_{\mu}$ of
the tangent space $T_{x}M$,
with basis vectors  labelled by $a = 0, 1, 2, 3$.
The space $F(M)$ of all frames (with arbitrary~$x$) is called the
\emph{frame bundle}, and we denote by $F_{x}(M)$
the set of frames with base point $x$.
Note that the group $\mathrm{Gl}(4,\R)$ of invertible $4\times 4$ matrices $A^{a}_{b}$
acts from the right on $F_{x}(M)$, sending $e_{x}$ to the frame $e'_{x} = e_{x}A$
with $e'^{\mu}_{a} = e^{\mu}_{b}A^{b}_{a}$.
This action is free and transitive;
any two frames $e_x$ and $e'_x$ over the same point $x$ are related by
$e'_x = e_x A$ for a unique matrix
$A^{a}_{b}$.

For a given Lorentzian metric $g$, the \emph{orthonormal frame bundle}
$O^{g}(M) \subset F(M)$ is the space of all orthonormal frames
$e^{\mu}_{a}$, satisfying
$g_{\mu\nu}e^{\mu}_{a}e^{\nu}_{b} = \eta_{ab}$.
Since two orthonormal frames $e_{x}$ and $e'_{x}$ over the same point
$x$ differ by a Lorentz transformation $\Lambda$,
$e'_{x} = e_{x}\Lambda$,
the Lorentz group
$\OO$ acts freely and transitively on the set $O^{g}_{x}(M) \subset F_{x}(M)$ of orthonormal
frames based at $x$.


Specifying a metric $g$ at $x$ is equivalent
to specifying the set $O^{g}_{x}(M)$ of orthonormal frames.
Since $O^{g}_{x}(M) \subset F_{x}(M)$ is an orbit under the action of the Lorentz group
$\OO$ on $F_{x}(M)$, specifying the metric at $x$ is equivalent to
picking a point in the \emph{orbit space} $\cR_{x}(M) = F_{x}(M)/\OO$.
This is
the set of equivalence classes $[e_x]$ of frames at $x$,
where two frames $e_x$ and $e'_x$ are deemed equivalent
if they differ by a Lorentz transformation $\Lambda$,
$e'_x = e_x \Lambda$.
We denote the bundle of all equivalence classes
$[e_x]$ (with arbitrary $x$) by $\cR(M)$.

To describe fermions in the presence of GR, it will be convenient to view
a metric $g$ on $M$ as a \emph{section} of
$\cR(M)$;
a smooth map $g \colon M \rightarrow \cR(M)$
that takes a point $x$ to an equivalence class $[e_{x}]$
of frames at $x$.
The configuration space\footnote{In
first order formalisms such as the Palatini approach, one
considers the bigger configuration space of
metrics $g$ together with a connection $\nabla$.
This amounts to replacing $\cR(M)$ by
$J^{1}F(M)/\OO$.}
of general relativity can thus be seen as the space
$\Gamma(\cR(M))$ of sections of the bundle
$\cR(M)$.

\section{Fermionic fields in a fixed background}\label{sec:FixedBackground}

We start by describing fermionic fields on $M$ in the presence of a fixed background
metric $g$.
In order to do this, a number of choices have to be made, especially if we
wish to keep track of the transformation behaviour of spinors
under parity and time reversal.

The local transformation behaviour is fixed by
choosing one out of the eight possible Pin groups $\mathrm{Pin}^{abc}$, together with a
(not necessarily $\C$-linear)
representation $V$
that extends
the spinor representation of $\mathrm{Spin}^{\uparrow}(3,1) \subset \mathrm{Pin}^{abc}$.
For example, $V$ consists of $n$ copies of $\C^4$ in the case of $n$ Dirac fermions,
and it consists of $m$ copies of $\C^2$ in the case of
$m$ Majorana fermions\footnote{The requirement that $V$ extends to an $\R$-linear $\mathrm{Pin}^{abc}$-representation
may place restrictions on $a$, $b$ and $c$. For instance, the $\mathrm{Spin}^{\uparrow}(3,1)$-representation
$V = \C^2$ (a single Majorana fermion) extends to $\Pin^{abc}$
if and only if $a=1$ and $b=-1$. To be consistent with the topological restrictions derived in
\S\ref{sec:coveringgroups}, we therefore need $m\geq 2$.}.

Once a Pin group has been selected, the second choice one has to make is a choice of
\emph{Pin structure}.
A Pin structure is a twofold cover
$u \colon Q^{g} \rightarrow O^{g}(M)$ of the orthonormal frame bundle,
equipped  with a $\mathrm{Pin}^{abc}$-action that is compatible with the action of the
Lorentz group on $O^{g}(M)$.
The compatibility entails that
if $\tilde{\Lambda} \in \mathrm{Pin}^{abc}$
covers $\Lambda \in \OO$, then $u(q_{x}\tilde{\Lambda}) = u(q_{x})\Lambda$ for all
pin frames
$q_{x}$ in $Q^{g}$.
A pin frame $q_{x}$ is based at the same point
as its image, the frame $u(q_{x})$. We denote by $Q^{g}_{x}$ the set of pin frames
based at $x$.

For a given manifold $M$ and a given Pin group $\mathrm{Pin}^{abc}$,
a Pin structure may or may not exist, and if it does, it need not be unique.
The obstruction theory for this problem
has been completely solved for the Cliffordian Pin groups in \cite{Ka68},
and for the general case in \cite{Ch94}.

Once a Pin structure $Q^{g}$ has been chosen, one can construct the
associated bundle $S^{g} = (Q^{g} \times V) / \mathrm{Pin}^{abc}$
of \emph{spinors}.
A spinor $\psi_{x} = [q_{x},\vec{v}]$ at $x$
is thus an equivalence class of a pin frame $q_{x} \in Q^{g}_{x}$
and a vector $\vec{v} \in V$, where $(q_{x}\tilde{\Lambda},\vec{v})$
is identified with
$(q_{x},\tilde{\Lambda} \vec{v})$ for
every element $\tilde{\Lambda}$ of the Pin group $\mathrm{Pin}^{abc}$.

For a given background metric $g$, the
fermionic fields are then described by sections of the spinor bundle
$S^{g}$, that is, by smooth maps
$\psi \colon M \rightarrow S^{g}$ that assign to each space-time point $x$
a spinor $\psi_x$ based at $x$.
The configuration space for the fermionic fields at a fixed metric
$g$ is thus
the space $\Gamma(S^{g})$ of sections of the spinor bundle
$S^{g}$.

\section{Fermionic fields coupled to GR}\label{sec:FermionicGR}

We now wish to describe the configuration space for fermionic fields
coupled to gravity.
This is \emph{not} simply the product of the configuration space of general relativity
and that of a fermionic field; the main difficulty here
is that the very space $S^{g}$ where the spinor field
$\psi$ takes values
depends on the metric $g$.
A solution to this problem was proposed in \cite{BG92,AWW16} for the Riemannian case,
and in \cite{BaerGauduchonMoroianu2005,J16,J17,MuellerNowaczyk2017} for metrics of
Lorentzian signature.
In order to handle reflections, we need to adapt this procedure as follows.

First, we choose a twofold cover of $\mathrm{Gl}(4,\R)$ that
agrees with the universal cover $\widetilde{\mathrm{Gl}}_{+}(4,\R)$ over
$\mathrm{Gl}_{+}(4,\R)$.
In \S\ref{sec:coveringgroups} we show that there are only \emph{two} such covers, which, for want of a better name,
we will call $\mathrm{Gin}^{+}$ and $\mathrm{Gin}^{-}$.
Having made our choice of $\mathrm{Gin}^{\pm}$, we choose what one may call
a \emph{Gin structure};
a twofold cover $u \colon \hat{Q} \rightarrow F(M)$ with a
$\mathrm{Gin}^{\pm}$-action that is compatible with the $\mathrm{Gl}(4,\R)$-action
on $F(M)$.
Corresponding to every (not necessarily orthogonal) frame $e_x$,
there are thus two gin frames $\hat{q}_{x}$ and $\hat{q}'_{x}$.
If $\widetilde{A} \in \mathrm{Gin}^{\pm}$ covers $A \in \mathrm{Gl}(4,\R)$,
then the two gin frames corresponding to $e_x A$ are
$\hat{q}_{x}\widetilde{A}$ and $\hat{q}'_{x}\widetilde{A}$.

We denote by $\Pin^{\pm}$ the twofold cover of $\mathrm{O}(3,1)$ inside $\mathrm{Gin}^{\pm}$.
Choosing a Gin structure $\hat{Q}$ for the group $\mathrm{Gin}^{\pm}$ is
equivalent to
choosing a Pin structure $Q^{g}$ for the group $\Pin^{\pm}$.
Indeed, for every $\mathrm{Gin}^{\pm}$ structure $\hat{Q}$,
the preimage ${Q^g \subset \hat{Q}}$ of $O^{g}(M) \subset F(M)$ under the map $u \colon \hat{Q} \rightarrow F(M)$
is a $\mathrm{Pin}^{\pm}$-structure, since the restriction ${u^g \colon Q^g \rightarrow O^{g}(M)}$ of $u$ to $Q^g$
intertwines the $\Pin^{\pm}$-action on $Q^g$ with the action of the Lorentz group $\mathrm{O}(3,1)$ on $O^g(M)$.
Conversely, every
$\mathrm{Pin}^{\pm}$-structure $u \colon Q^{g} \rightarrow O^{g}(M)$
gives rise to the associated $\mathrm{Gin}^{\pm}$-structure
$\hat{Q} = (Q^{g}\times \mathrm{Gin}^{\pm})/\Pin^{\pm}$. This is the
space of equivalence classes $[q_{x},\tilde{A}]$, where
$(q_{x}\tilde{\Lambda},\tilde{A})$ is identified with
$(q_{x}, \tilde{\Lambda}\tilde{A})$
for every $\tilde{\Lambda}$ in $\Pin^{\pm}$.
The obstruction theory for $\mathrm{Gin}^{\pm}$-structures therefore reduces to
the obstruction theory for $\mathrm{Pin}^{\pm}$-structures, which has been
worked out in \cite{Ch94}.

In analogy with \cite{MuellerNowaczyk2017}, one constructs
the \emph{universal spinor bundle}
$\cE = {(\hat{Q} \times V)/ \mathrm{Pin}^{\pm}}$ using the Gin structure $\hat{Q}$.
A universal spinor $\Psi_{x} = [\hat{q}_{x},\vec{v}]$ at $x$
is an equivalence class of a gin frame $\hat{q}_{x} \in \hat{Q}_{x}$
and a vector $\vec{v} \in V$, where $(\hat{q}_{x} \tilde{\Lambda},\vec{v})$
is identified with
$(\hat{q}_{x},\tilde{\Lambda} \vec{v})$ for
every $\tilde{\Lambda}$ in $\mathrm{Pin}^{\pm}$.
Note that a universal spinor $\Psi_{x}$
in $\cE = (\hat{Q} \times V)/ \mathrm{Pin}^{\pm}$
defines a metric $g_{\mu\nu}$ at $x$, together with
a spinor
$\psi_{x}$ in the spinor bundle
$S^{g} = (Q^{g} \times V) / \mathrm{Pin}^{\pm}$
\emph{that corresponds with the metric~$g_{\mu\nu}$ induced by $\Psi_{x}$}.

Indeed, since the covering map $u \colon \hat{Q} \rightarrow F(M)$
intertwines the $\mathrm{Pin}^{\pm}$-action on $\hat{Q}$ with the
$\mathrm{Gl}(4,\R)$-action on $F(M)$, it identifies
the quotient of $\hat{Q}$ by $\mathrm{Pin}^{\pm}$
with
the quotient of $F(M)$ by $\OO$, which is the orbit space $\cR(M)$.
From a universal spinor $\Psi_{x} = [\hat{q}_{x},\vec{v}]$ at~$x$, we thus
obtain an equivalence class $[u(\hat{q}_{x})]$ in $\cR_{x}(M)$, and hence
a metric~$g_{\mu\nu}$ at the point $x$.

To obtain not only the metric $g_{\mu\nu}$ but also the spinor $\psi_{x}$,
recall that the Pin structure $Q^{g}$
corresponding to $g_{\mu\nu}$ is the preimage of $O^{g}(M)$
under the double cover ${u \colon \hat{Q} \rightarrow F(M)}$.
Since $Q^{g} \subset \hat{Q}$ contains the gin frame
$\hat{q}_{x}$,
the equivalence class $\Psi_{x} = [\hat{q}_{x},\vec{v}]$
in $\cE = (\hat{Q} \times V)/ \mathrm{Pin}^{\pm}$ yields
an equivalence class $\psi_{x} = [q_{x},\vec{v}]$ in the spinor bundle
$S^{g} = (Q^{g} \times V) / \mathrm{Pin}^{\pm}$ by setting $q_{x} = \hat{q}_{x}$.
Here, $S^{g}$ is the spinor bundle derived
from the metric $g_{\mu\nu}$ that is induced by $\Psi$.

We conclude that \emph{both} the
metric $g$ \emph{and} the fermionic field $\psi$
are described by a \emph{single}
section $\Psi \colon M \rightarrow \cE$, a smooth map
assigning to each point $x$ of space-time a universal spinor $\Psi_{x}$ based at $x$.
The configuration space of fermionic fields coupled to gravity
is thus the space $\Gamma(\cE)$ of sections of the universal spinor
bundle $\cE$.

\section{Covering groups}\label{sec:coveringgroups}

Out of the eight Pin groups covering $\OO$,
the only two that are compatible with this formalism
are the twofold cover $\mathrm{Pin}^{+}$
of $\OO$ inside $\mathrm{Gin}^{+}$, and the
twofold cover $\mathrm{Pin}^{-}$
of $\OO$ inside $\mathrm{Gin}^{-}$.
We show that their
coefficients in the sense of \S\ref{sec:Introduction}
are $(a,b,c) = (+,+,-)$ and $(a,b,c) = (-,-,-)$.

First we show that there are only two double covers of $\mathrm{Gl}(4,\R)$ that reduce to the universal cover
over $\mathrm{Gl}_{+}(4,\R)$. Assume that $G$ is such a cover. If
$\Lambda_{T}$ is an element of $G$ that covers the time reversal operator
$T \in \mathrm{Gl}(4,\R)$, then the automorphism
$\mathrm{Ad}_{\Lambda_T}(\widetilde{A}):= \Lambda_T \widetilde{A} \Lambda_{T}^{-1}$
of $\widetilde{\mathrm{Gl}}_{+}(4,\R)$ covers the automorphism $\mathrm{Ad}_{T}(A) := TAT^{-1}$ of $\mathrm{Gl}_{+}(4,\R)$.
By the universal covering property,
$\Lambda_T \widetilde{A} \Lambda_{T}^{-1}$ is uniquely determined by $\widetilde{A}$, and it
depends neither on
the choice of $G$, nor on the choice of $\Lambda_{T}$ inside $G$.
Since every element of $G$ can be written as either $\widetilde{A}$ or $\widetilde{B}\Lambda_{T}$,
there are four types of products, namely those of the form $\widetilde{A}\widetilde{A}'$,
$\widetilde{A}(\widetilde{B}\Lambda_{T})$, $(\widetilde{B}\Lambda_{T})\widetilde{A}$ and
$(\widetilde{B}\Lambda_{T})(\widetilde{B}'\Lambda_{T})$, where
$\widetilde{A}, \widetilde{A}', \widetilde{B}, \widetilde{B}'$ are in $\widetilde{\mathrm{Gl}}_{+}(4,\R)$.
Products of the first 2 types are determined by the group structure on
$\widetilde{\mathrm{Gl}}_{+}(4,\R)$. This is true for the third type as well, since
$(\widetilde{B}\Lambda_{T})\widetilde{A} = \widetilde{B}(\Lambda_{T}\widetilde{A}\Lambda_{T}^{-1})\Lambda_{T}$,
and $\Lambda_{T}\widetilde{A}\Lambda_{T}^{-1}$ is independent of $G$.
As $(\widetilde{B}\Lambda_{T})(\widetilde{B}'\Lambda_{T}) = (\widetilde{B}(\Lambda_{T}\widetilde{B}'\Lambda_{T}^{-1}))\Lambda^2_{T}$, the only choice in the product structure on $G$ lies in the sign of
$\Lambda_{T}^{2} = \pm 1$, yielding the two groups $\mathrm{Gin}^{\pm}$.
The twofold cover $\mathrm{Pin}^{+}$ of $\OO$ inside $\mathrm{Gin}^{+}$ thus has $b = +1$, whereas the
twofold cover $\mathrm{Pin}^{-}$ inside $\mathrm{Gin}^{-}$ has $b=-1$.

To establish that both $\Pin^{+}$ and $\Pin^{-}$ satisfy $c=-1$, note that although
the central element $PT = \mathrm{diag}(-1,-1,-1,-1)$
does not lie in the connected component of unity for the Lorentz group
$\OO$, it does lie in the
connected subgroup $\mathrm{SO}(4)$ of $\mathrm{Gl}_{+}(4,\R)$.
As the inverse image of $\mathrm{SO}(4)$ under the universal cover
$\widetilde{\mathrm{Gl}}_{+}(4,\R) \rightarrow \mathrm{Gl}_{+}(4,\R)$ is
its universal cover $\mathrm{Spin}^{\uparrow}(4)$,
the square of $\Lambda_{P}\Lambda_{T}$ inside
$\widetilde{\mathrm{Gl}}_{+}(4,\R)$
equals its square in $\mathrm{Spin}^{\uparrow}(4)$.
Here, the elements
$\pm i\gamma_5 = \mp \gamma_0 \gamma_1 \gamma_2 \gamma_3$
that cover $PT$ square to $+1$, as one easily derives using
the Clifford relations $\{\gamma_\mu, \gamma_{\nu}\} = 2\delta_{\mu\nu}$
for the Euclidean gamma matrices $\gamma_{\mu}$.
It follows that $(\Lambda_P\Lambda_T)^2 = 1$, and hence $c=-1$.

It remains to show that $a=b$. For this, note that
the restriction of the automorphism
$\mathrm{Ad}_{\Lambda_{T}}$ of $\widetilde{\mathrm{Gl}}_{+}(4,\R)$
to the simply connected subgroup $\mathrm{Spin}^{\uparrow}(4) \subset
\widetilde{\mathrm{Gl}}_{+}(4,\R)$
is uniquely determined by its induced Lie algebra automorphism.
On $\mathrm{Spin}^{\uparrow}(4)$, we thus have
$\mathrm{Ad}_{\Lambda_{T}}(u) = \gamma_0 u \gamma_0^{-1}$.
As $\gamma_0 (i\gamma_5)\gamma^{-1}_0 = -i\gamma_5$, we find that
$\Lambda_{T}\Lambda_{P} = \mathrm{Ad}_{\Lambda_{T}}(\Lambda_{P}\Lambda_{T})
= -\Lambda_{P}\Lambda_{T}$.
As we already established that $(\Lambda_{P}\Lambda_{T})^2=1$,
it follows that $\Lambda^2_{P} \Lambda_{T}^2 = - 1$, and hence that
$a = -\Lambda_{P}^2$ is equal to $b = \Lambda^2_{T}$.
We thus conclude that $(a,b,c) = (+,+,-)$ for $\Pin^{+}$, and $(a,b,c) = (-,-,-)$ for $\Pin^-$.

The groups $\Pin^{+}$ and $\Pin^{-}$ are therefore \emph{not} isomorphic to
the Cliffordian Pin groups
$\Pin(3,1)$ and $\Pin(1,3)$. These are generated by the Clifford elements $v^{\mu}\tilde{\gamma}_{\mu}$
with $\eta_{\mu\nu}v^{\mu}v^{\nu} = \pm 1$, where
the Lorentzian gamma matrices $\tilde{\gamma}_{\mu}$ satisfy
$\{\tilde{\gamma}_\mu, \tilde{\gamma}_{\nu}\} = 2\eta_{\mu\nu}$ for $\Pin(3,1)$, and
$\{\tilde{\gamma}_\mu, \tilde{\gamma}_{\nu}\} = -2\eta_{\mu\nu}$ for $\Pin(1,3)$.
Since the group elements covering $P$ and $T$ are
$\Lambda_P = \tilde{\gamma}_1\tilde{\gamma}_2\tilde{\gamma}_3$ and
$\Lambda_T = \tilde{\gamma}_0$,
one readily verifies that
$(a,b,c) = (+,-,+)$ for $\Pin(3,1)$, and that
 $(a,b,c) = (-,+,+)$
for $\Pin(1,3)$ (cf.~\cite{Da88,Ch94}).

In particular, we conclude that
the two Pin groups $\Pin^{\pm}$ compatible with GR
are \emph{not} the  widely used Cliffordian Pin groups
$\Pin(3,1)$
and $\Pin(1,3)$.

\section{Transformation under diffeomorphisms}\label{sec:infdiffeo}

In the above derivation of the two admissible Pin groups, a crucial role is played by the
continuous covering map $u \colon \hat{Q} \rightarrow F(M)$.
This map has physical significance, since it
induces an
infinitesimal action of the space-time diffeomorphism group $\mathrm{Diff}(M)$
on the configuration space of fermions coupled to gravity (cf.~\cite{AWW16,J17}).
This allows one to formulate a
theory which is (up to sign) covariant under general coordinate transformations (cf. \cite{DD13,AWW16}),
and to construct a Stress-Energy-Momentum tensor via Noether's theorem
(cf.~\cite{GM92,FR04}, and cf.~\cite[\S6]{BaerGauduchonMoroianu2005} for an approach
using variation of the metric).

To construct the infinitesimal action, note that $\mathrm{Diff}(M)$ acts by automorphisms on the frame bundle $F(M)$,
a diffeomorphism $\phi$ maps $e_x \in F_{x}(M)$ to
$D\phi (e_x):= \partial_{\overline{\mu}}\phi^{\mu}e^{\overline{\mu}}_{a}$ in $F_{\phi(x)}(M)$.
A one-parameter group $\phi_{\varepsilon}$ of diffeomorphisms thus yields
a one-parameter group $D\phi_{\varepsilon}$ of automorphisms
of $F(M)$.
Since $u \colon \hat{Q} \rightarrow F(M)$
is a double cover, this lifts to a
unique one-parameter group $D\hat{\phi}_{\varepsilon}$ of automorphisms
of $\hat{Q}$.
On the universal spinor bundle $\Sigma = (\hat{Q} \times V)/\Pin^{\pm}$,
we define the lift by
$
D\hat{\phi}_{\varepsilon}[\hat{q}_{x},\vec{v}] =
[D\hat{\phi}_{\varepsilon}(\hat{q}_{x}),\vec{v}]
$.
For the infinitesimal variation of the universal spinor field
$\Psi \colon M \rightarrow \Sigma$ along $\phi_{\varepsilon}$,
this yields
$\delta \Psi_{x} = \frac{d}{d \varepsilon}|_{0}
D\hat{\phi}_{\varepsilon}(\Psi_{\phi^{-1}_{\varepsilon}(x)})$.

\section{The role of diffeomorphisms in restricting the Pin groups}

We stress that the above restrictions on the Pin groups
are not needed to \emph{construct} the configuration space for
fermions coupled to gravity, but to ensure that it \emph{transforms} appropriately
under space-time diffeomorphisms.

Indeed, to construct the configuration space,
one could simply
choose any principal $\Pin^{abc}$-bundle
$P \rightarrow \cR(M)$ (for example the trivial one), and
construct the universal spinor bundle
$\Sigma = (P \times V) / \Pin^{abc}$
as in \S\ref{sec:FermionicGR}. Its sections $\Psi \in \Gamma(\Sigma)$
can be interpreted as a fermionic field $\psi$ together with a metric~$g$,
so $\Gamma(\Sigma)$ may serve as a configuration space.
This requires no restrictions on the Pin groups,
nor on the topology of $M$.

However, this simple construction leaves the space-time transformation behaviour undetermined.
We show that the restrictions on the Pin groups are recovered by imposing appropriate transformation
behaviour on $\Gamma(\Sigma)$.
Compatibility with the Lorentz group leads to the familiar restrictions on the topology of $M$,
compatibility with infinitesimal diffeomorphisms leads to Pin groups with $c=-1$, and
compatibility with a double cover of the diffeomorphism group requires Pin groups with $a=b$ as well as  $c=-1$.

\subsection{Lorentz transformations}\label{subsecLorentz}

The \emph{pullback} of a bundle $E \rightarrow Y$ along a map $f \colon X \rightarrow Y$
is the bundle $f^*E \rightarrow X$ with $(f^*E)_{x} := E_{f(x)}$.
Starting from the principal bundle $P \rightarrow \cR(M)$, one thus obtains
for every metric $g \colon M \rightarrow \cR(M)$ a principal $\Pin^{abc}$-bundle $g^*P \rightarrow M$.
Its fibre $g^*P_{x}$ at $x$ is
the fibre $P_{g(x)}$ of $P$ at $g(x)\in \cR(M)$.
The bundle $g^*P$ is not quite a Pin structure, since the action of
$\Pin^{abc}$ on $g^*P$ is as yet unrelated to the action of $\mathrm{O}(3,1)$ on $O^{g}(M)$.
To define the transformation behaviour of $\Psi$ under infinitesimal isometries,
we need to choose a Pin structure on each of the bundles~$g^*P$. That is,
for any possible metric
$g\in \Gamma(\cR(M))$,
we need to choose a
double cover $u^{g} \colon g^*P \rightarrow O^{g}(M)$ that intertwines the action of $\Pin^{abc}$ on $g^*P$
with the action of $\mathrm{O}(3,1)$ on $O^g(M)$.
This is where the restrictions on the topology of $M$ arise:
if the conditions in \cite{Ch94} are met, then
it is possible to endow every \emph{single} bundle $g^*P \rightarrow M$ with a double covering map
$u^{g} \colon g^*P \rightarrow O^{g}(M)$, making it into a Pin structure.

\subsection{Infinitesimal diffeomorphisms}\label{subsec:InfDiffeo}

The problem is that, in general, these covering maps $u^{g}$ do not
depend continuously on the metric $g$.
If we require this to be the case, then we recover
the infinitesimal action of the diffeomorphism group on the configuration space,
as well as the restriction $c=-1$ on the Pin groups.
This already excludes the `Cliffordian' Pin groups $\Pin(3,1)$ and $\Pin(1,3)$.

If we pull back $P \rightarrow \cR(M)$ along the evaluation
map $\mathrm{ev} \colon {M \times \Gamma(\cR(M))} \rightarrow \cR(M)$, defined as $\mathrm{ev}(x,g) := g(x)$,
we obtain the principal $\Pin^{abc}$-bundle
$\mathrm{ev}^*P \rightarrow {M\times \Gamma(\cR(M))}$.
It consists of all pairs
$(p,g) \in P\times \Gamma(\cR(M))$
where $p$ lies in $g^*P$.
The maps $u^g$ for the different metrics $g\in \Gamma(\cR(M))$ then combine to a single
map $u \colon \mathrm{ev}^*P \rightarrow F(M)$, defined by $u(p,g) := u^g(p)$.
We say that $u^{g}$ \emph{depends continuously on} $g$ if the map $u \colon \mathrm{ev}^*P \rightarrow F(M)$
is continuous.


If $u^g$ depends continuously on $g$, then we obtain an infinitesimal action of
$\mathrm{Diff}(M)$ on the configuration space $\Gamma(\Sigma)$ of fermions
coupled to gravity.
Since the (left) action of $\mathrm{Diff}(M)$ on $F(M)$ commutes with the (right)
action of $\mathrm{Gl}(4,\R)$, we have an action of $\mathrm{Diff}(M)$ on $\cR(M)$, yielding
the usual space-time transformation behaviour
$g_x \mapsto D\phi \,g_{\phi^{-1}(x)}$ on the space $\Gamma(\cR(M))$ of metrics.
 To obtain the transformation behaviour of spinors coupled to gravity,
note that since $u \colon \mathrm{ev}^*P \rightarrow F(M)$ is continuous, it induces a double cover
from $\mathrm{ev}^*P$ to $\mathrm{ev}^*F(M)$, the space of all pairs $(e_x, g) \in F(M)\times \Gamma(\cR(M))$
with $e_x \in O^g(M)$. Since $\mathrm{Diff}(M)$ acts on $\mathrm{ev}^*F(M)$,
it has an \emph{infinitesimal} action on the double cover $\mathrm{ev}^*P$. This yields an infinitesimal action on
$\mathrm{ev}^*\Sigma \rightarrow M \times \Gamma(\cR(M))$, the space of all pairs $([\hat{q}_x,v],g) \in \Sigma \times \Gamma(\cR(M))$ where $\hat{q}_x$ is in $g^*P$. This yields an infinitesimal action on $\Gamma(\Sigma)$, since
a section $\Psi \in \Gamma(\Sigma)$ can be viewed as a map from
$M$ to $\mathrm{ev}^*\Sigma$, sending $x\in M$ to the pair $(\Psi_{x},g)$, where $g$ is the
metric obtained from the section $\Psi$.

To recover the restriction $c=-1$,
consider the case $M=\R^4$.
Since $\mathrm{Gl}(4,\R)$ is a subgroup of $\mathrm{Diff}(\R^4)$, it acts from the left on $F(\R^4)$, and hence on $\cR(\R^4)$.
Since only the Lorentz group $\mathrm{O}(3,1)$ leaves the Minkowski metric $\eta$ invariant,
we obtain
an injective, continuous map
$\sigma \colon \mathrm{Gl}(4,\R)/\mathrm{O}(3,1) \rightarrow \R^4 \times \Gamma(\cR(\R^4))$ by
$\sigma([A]) := (0,A\eta)$.
The pullback bundle $\sigma^*\mathrm{ev}^*P$ is a principal
$\Pin^{abc}$-bundle over $\mathrm{Gl}(4,\R)/\mathrm{O}(3,1)$.
Note that $\mathrm{ev}\circ \sigma$ is a diffeomorphism from $\mathrm{Gl}(4,\R)/\mathrm{O}(3,1)$
to $\cR_{0}(\R^4)$, the space of all Lorentzian metrics on the tangent space $T_{0}\R^4$
at the origin, so
$\sigma^*\mathrm{ev}^*P = (\mathrm{ev} \circ \sigma)^*P$ can be identified with the restriction
$P_0$
of $P$ to $\cR_{0}(\R^4)$.
Since the image of the pullback map
$\sigma^*u \colon \sigma^*\mathrm{ev}^*P \rightarrow F(\R^4)$
is the set $F_0(\R^4) \simeq \mathrm{Gl}(4,\R)$ of frames at the origin, we obtain a continuous double cover
$P_{0} \rightarrow \mathrm{Gl}(4,\R)$.
As this double cover intertwines the (right)
$\Pin^{abc}$-action on $P_0$ with the (right) $\mathrm{O}(3,1)$-action on $F_0(\R^4) \simeq \mathrm{Gl}(4,\R)$,
the preimage $P_0^{+}$ of $\mathrm{Gl}(4,\R)^{+}$ is the universal covering group
$\widetilde{\mathrm{Gl}}(4,\R)^{+}$, and the orientation-preserving subgroup of $\mathrm{Pin}^{abc}$
coincides with the subgroup of $\widetilde{\mathrm{Gl}}(4,\R)^{+}$ that covers $\mathrm{SO}(3,1)$.
Since $(\Lambda_{P}\Lambda_{T})^2 = 1$ in $\widetilde{\mathrm{Gl}}(4,\R)^{+}$,
we recover the restriction $c=-1$ of
\S\ref{sec:coveringgroups}.


\subsection{Double cover of the diffeomorphism group}\label{subsec:DoubleDiffeo}

In the above line of reasoning, the group structure on $P_{0}^{+}$ stems from its identification with the universal cover
of the connected Lie group $\mathrm{Gl}(4,\R)^{+}$. Since we lack a group structure on the disconnected space $P_0$,
we cannot directly infer that $a=b$.
This does, however, follow from the slightly stronger assumption
that the $\mathrm{Diff}(M)$-action on $\mathrm{ev}^*F(M)$
lifts to an action by automorphisms of a double cover $\mathrm{\widehat{Diff}}(M)$ on $\mathrm{ev}^*P$.
This yields an action of $\widehat{\mathrm{Diff}}(M)$ on $\mathrm{ev}^*\Sigma$, and by
by identifying $\Psi \in \Gamma(\Sigma)$
with a map from $M$ to $\mathrm{ev}^*\Sigma$ as before, one obtains an action of
$\widehat{\mathrm{Diff}}(M)$ on $\Gamma(\Sigma)$.
Explicitly, $\phi\in \mathrm{Diff}(M)$ acts on $\mathrm{ev}^*F(M)$ by taking $(e_x,g)$ to $(D\phi (e_x), D\phi \circ g \circ \phi^{-1})$. If this lifts to an automorphism $D\hat{\phi}$ of
$\mathrm{ev}^*\Sigma$,
then $D\hat{\phi}$ maps $\Psi \in \Gamma(\Sigma)$ to the unique $\Psi' \in \Gamma(\Sigma)$ with
$(\Psi'_{x}, D\phi \circ g \circ \phi^{-1}) = D\hat{\phi} (\Psi_{\phi^{-1}(x)}, g)$.

To see that this yields the restriction $a=b$, consider the case $M=\R^4$. Then
$\mathrm{Gl}(4,\R)$ is a subgroup of $\mathrm{Diff}(\R^4)$, and its preimage in $\widehat{\mathrm{Diff}}(\R^4)$
is one of the two Gin groups
$\mathrm{Gin}^{\pm}$.
The (left) action of $\mathrm{Gin}^{\pm}$ by automorphisms on $\mathrm{ev}^*P$
covers the (left) action of $\mathrm{Gl}(4,\R)$ by automorphisms
on $\mathrm{ev}^*F(M)$, so
in particular, the (left) action of $\mathrm{Gin}^{\pm}$ on $\sigma^*\mathrm{ev}^*P = P_0$ covers the
(left) action of $\mathrm{Gl}(4,\R)$ on $\sigma^*\mathrm{ev}^*F(\R^4) = F_{0}(\R^4)$.
This intertwines the (right) action of $\Pin^{abc}$ on $P_0$ with the (right) action of
$\mathrm{O}(3,1)$ on $F_0(\R^4)$. Since all these actions are free,
we can identify $\Pin^{abc}$ with a subgroup of $\mathrm{Gin}^{\pm}$ that covers the Lorentz group
$\mathrm{O}(3,1)$.
Following the line of reasoning in \S\ref{sec:coveringgroups}, we thus find $a=b$  as well as
$c = -1$.

We conclude that although an infinitesimal action of the space-time diffeomorphism group on the configuration space of fermions coupled to gravity requires $c=-1$, an action of a double cover of the diffeomorphism group can only be achieved if the Pin group additionally satisfies the relation $a=b$.

\section{Discussion}

The conclusion that
only two of the eight Pin groups
are compatible with general relativity,
appears to be quite robust.
It is based on the elementary observation
that the twofold spin cover of the orthonormal frame bundle $O^{g}(M)$
is compatible with a twofold cover
of the \emph{full} frame bundle $F(M)$.
Although we derived this from the setting
outlined in \S\ref{sec:FermionicGR} (going back to \cite{Ko72,BG92,AWW16}
in the Riemannian and \cite{BaerGauduchonMoroianu2005,J16,J17,MuellerNowaczyk2017} in the Lorentzian case),
the use of double covers of the full frame bundle
-- and hence our conclusion that only two Pin groups are admissible --
is common to many other approaches, such as the more
`global' formalism developped in \cite{DP86,Da88,Swift93,DD13}.
In fact, the restrictions on the Pin groups are closely linked to the transformation behaviour of fermions
coupled to gravity under space-time diffeomorphisms.



Since \emph{any} principal bundle
with an infinitesimal action of the
space-time diffeomorphism group is associated to
a discrete cover of a (higher order) frame bundle~\cite{J16,J17},
we expect that
our restrictions on the Pin group are not
an artefact of the particular
description that we have adopted.

\subsection*{Acknowledgments}
I would like to thank Edward Witten for several valuable comments.
This research is supported by the NWO grant 639.032.734
``Cohomology and representation theory of infinite dimensional Lie groups''.

\end{document}